\documentclass[review]{elsarticle}

\usepackage{hyperref}

\journal{Journal of \LaTeX\ Templates}

\begin{document}

\begin{frontmatter}

\title{Global constraints on $Z_2$ fluxes in two different anisotropic limits of a hypernonagon Kitaev model}


\author[address1]{Yasuyuki~Kato}
\author[address2]{Yoshitomo~Kamiya}
\author[address3]{Joji~Nasu}
\author[address1]{Yukitoshi~Motome}
%
\address[address1]{Department of Applied Physics, University of Tokyo, Hongo, 7-3-1, Bunkyo, Tokyo 113-8656, Japan}
\address[address2]{Condensed Matter Theory Laboratory, RIKEN, Wako, Saitama 351-0198, Japan}
\address[address3]{Department of Physics, Tokyo Institute of Technology, Meguro, Tokyo 152-8551, Japan}

\begin{abstract}
The Kitaev model is an exactly-soluble quantum spin model, whose ground state provides a canonical example of a quantum spin liquid. 
Spin excitations from the ground state are fractionalized into emergent matter fermions and $Z_2$ fluxes. 
The $Z_2$ flux excitation is pointlike in two dimensions, while it comprises a closed loop in three dimensions because of the local constraint for each closed volume. 
In addition, the fluxes obey global constraints involving (semi)macroscopic number of fluxes. 
We here investigate such global constraints in the Kitaev model on a three-dimensional lattice composed of nine-site elementary loops, 
dubbed the hypernonagon lattice, whose ground state is a chiral spin liquid.
We consider two different anisotropic limits of the hypernonagon Kitaev model where 
the low-energy effective models are described solely by the $Z_2$ fluxes. 
We show that there are two kinds of global constraints in the model defined on a three-dimensional torus, namely,
surface and volume constraints: 
the surface constraint is imposed on 
the even-odd parity of the total number of fluxes threading
a two-dimensional slice of the system, 
while the volume constraint is for the even-odd parity of the number of
the fluxes through specific plaquettes whose total number is proportional to the system volume. 
In the two anisotropic limits, therefore, the elementary excitation of $Z_2$ fluxes occurs in a pair of closed loops so as to satisfy both two global constraints as well as the local constraints. 
\end{abstract}


\end{frontmatter}


\section{Introduction}
The Kitaev model is an exactly-soluble quantum spin model despite the severe frustration~\cite{kitaev2006}. 
Since the ground state was rigorously proven to be a quantum spin liquid, this model has been extensively studied in this decade, 
not only on a honeycomb lattice in the original proposal~\cite{kitaev2006} but also on many tricoordinate lattices in both two and three dimensions~\cite{yao2007,si2008,mandal2009,hermanns2015,obrien2016}. 
In the ground state, the quantum spins are fractionalized into two types of emergent quasiparticles, matter fermions and $Z_2$ fluxes~\cite{kitaev2006}. 
The $Z_2$ flux is a static conserved quantity 
defined for each elementary loop of the lattice. 
Thus, the eigenstates are classified into the sectors with different configurations of the $Z_2$ fluxes. 
For example, the ground state of the honeycomb Kitaev model is obtained in the zero-flux sector where all the fluxes are +1~\cite{kitaev2006}. 
In an anisotropic limit of the exchange interactions, the matter fermion excitation is gapped away, 
and the low-energy effective Hamiltonian is described solely by the $Z_2$ fluxes~\cite{kitaev2003}.

The $Z_2$ flux excitation is local and pointlike in the two-dimensional Kitaev models, whereas it forms a closed loop in three dimensions~\cite{mandal2009,obrien2016,kimchi2014}. 
This is due to the local constraints arising from the algebra of Pauli matrices on any closed volume in the lattice. 
In addition, there are some global constraints on the $Z_2$ fluxes. 
For instance, in the anisotropic limit of the hyperhoneycomb lattice, two different types of the global constraints were discussed: surface and volume constraints~\cite{mandal2014}.
Thus, the elementary excitation of the $Z_2$ fluxes in the three-dimensional systems may acquire peculiar nature because of both the local and global constraints. 
Indeed, in the anisotropic hyperhoneycomb case, the lowest-energy excitation is a pair of smallest loops~\cite{mandal2014}.  

In this paper, we examine the constraints on the $Z_2$ fluxes
in the Kitaev model defined on another three-dimensional lattice, which we call the hypernonagon lattice 
(also known as (9,3)a in the classification of Wells~\cite{wells1977,obrien2016}).
The hypernonagon lattice is composed of nine-site elementary loops~(Fig.~\ref{lattice}). 
Odd-number loops accommodate the $Z_2$ fluxes that are odd under both time reversal and parity (spatial inversion) transformations, and in fact,
the ground state of the model is a chiral spin liquid with spontaneous breaking of 
both time reversal and spatial inversion symmetries~\cite{kitaev2006}. 
In our resent study~\cite{kato2017}, we derived the low-energy effective models in two different anisotropic limits of the hypernonagon Kitaev model, 
and investigated the finite-temperature phase transition to the chiral spin liquid by using the classical Monte Carlo simulation.
In the simulation, the Monte Carlo updates were performed by a simultaneous flip of a pair of closed loops, not to violate both the local and global loops. 
We here discuss the details of the global constraints that were not presented in the previous study.

\section{Kitaev model and local constraint}
\begin{figure}[!htb]
  \centering
  \includegraphics[width=\columnwidth,trim = 0 0 0 0,clip]{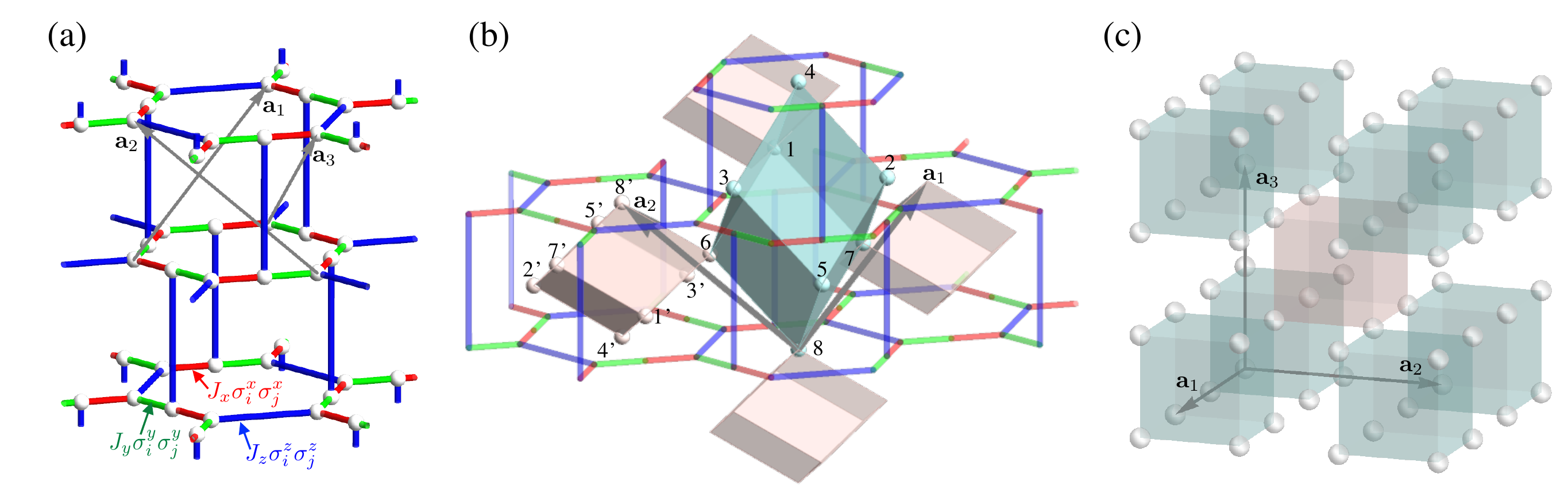}
  \caption{
    (a) The hypernonagon lattice also known as (9,3)a~\cite{wells1977,obrien2016}.
    The red, green, and blue bonds represent the $x$, $y$, and $z$ bonds in the Kitaev model in Eq.~(\ref{eq:Kitaev_Hamiltonian}), respectively.
    ${\bf a}_1$, ${\bf a}_2$, and ${\bf a}_3$ are the primitive vectors. 
    The white spheres represent the sites for $S=1/2$ quantum spins ($\sigma$ spins).
    (b) A  distorted cubic lattice where the spheres represent
    nine-site loops accommodating the local conserved quantities, $W_p$.
    (c) The simple cubic lattice topologically equivalent to (b). 
  }
  \label{lattice}
\end{figure}

The Hamiltonian of the Kitaev model on the hypernonagon lattice is given by 
\begin{eqnarray}
\mathcal{H}=
-J_x \sum_{\langle i,j \rangle_x} \sigma^x_i \sigma^x_j
-J_y \sum_{\langle i,j \rangle_y} \sigma^y_i \sigma^y_j
-J_z \sum_{\langle i,j \rangle_z} \sigma^z_i \sigma^z_j,
\label{eq:Kitaev_Hamiltonian}
\end{eqnarray}
where $\sigma_i^\mu$ is the $\mu$ component of the Pauli matrices representing a $S=1/2$ spin at site $i$.
The sum of $\langle i,j \rangle_\mu$ runs over the nearest-neighbor sites on all $\mu$ bonds of the hypernonagon lattice [see Fig.~\ref{lattice}(a)], 
and $J_\mu$ is the coupling constant for each type of bonds.
We consider the periodic boundary condition hereafter.

In the hypernonagon lattice, the unit cell includes eight elementary nine-site loops, 
which are represented by the eight corners of the bluish (B) cube in Fig.~\ref{lattice}(b).
Each nine-site loop accommodates a local conserved quantity, $W_p$.
Following Refs.~\cite{obrien2016,kato2017}, 
we define $W_p$ by a product of bond operators $\sigma_i^\mu\sigma_j^\mu$ for all bonds on the loop surrounding the plaquette $p$ 
in a clockwise manner viewed from the center of B cube in Fig.~\ref{lattice}(b): 
$W_p = \prod_{ \langle i,j \rangle_\mu \in p} \sigma^\mu_i \sigma^\mu_j$. 
Then, $W_p$ takes $\pm i$, which is called the $\pm \pi/2$ flux.
In each unit cell, there are two local constraints arising from the algebra of Pauli matrices: 
the product of all the $W_p$ in a B 
as well as that of all the $W_p$ in reddish (R) cube are both unity.

\section{Surface constraint}
\begin{figure}[!htb]
  \centering
  \includegraphics[width=0.75 \columnwidth,trim = 0 0 0 0,clip]{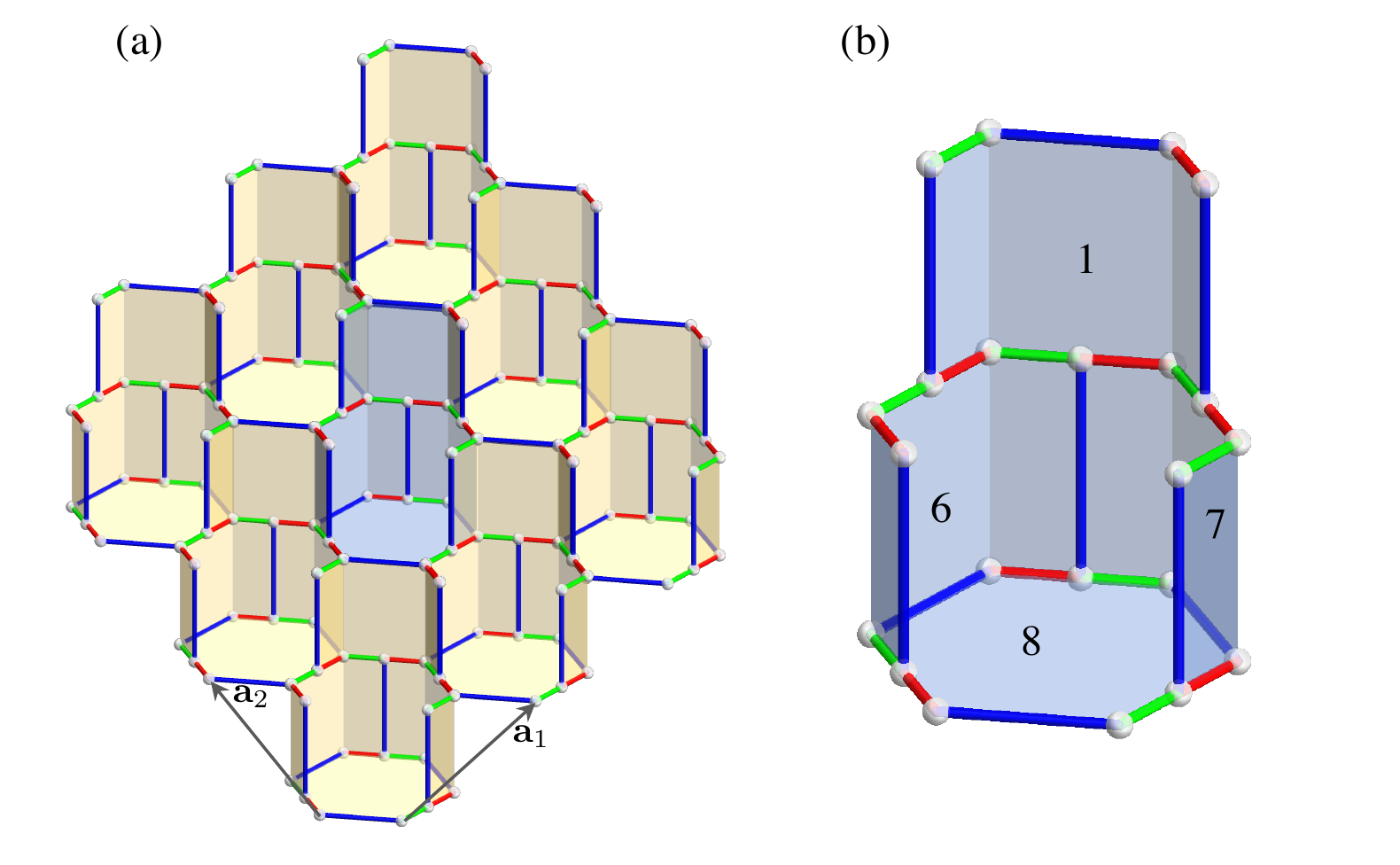}
  \caption{
  (a) Surface spanned by ${\bf a}_1$ and ${\bf a}_2$.
  (b) Building block of the surface.
  }
  \label{surface}
\end{figure}

Next, we consider a global constraint on $W_p$ imposed on a two-dimensional slice of the three-dimensional hypernonagon lattice.  
This is called the surface constraint.
Figure~\ref{surface}(a) shows an example of such surfaces, $\mathcal{S}$, spanned by ${\bf a}_1$ and ${\bf a}_2$.
It is noteworthy that this surface $\mathcal{S}$ corresponds to an ${\bf a}_1$-${\bf a}_2$ plane in the cubic lattice representation in Fig.~\ref{lattice}(c).
As shown in Fig.~\ref{surface}(b), the building block of $\mathcal{S}$ is a set of four $W_p$: $W_1$, $W_6$, $W_7$, and $W_8$ in each unit cell
[see Fig.~\ref{lattice}(b) for the sublattice numbering].
When taking the product of all $W_p$ on the surface $\mathcal{S}$, the algebraic properties of Pauli matrices lead to
\begin{equation}
\prod_{p \in \mathcal{S}} W_p =1.
\end{equation}
This identity is applied to any surfaces spanned by ${\bf a}_1$ and ${\bf a}_2$, and also to those spanned by ${\bf a}_1$ and ${\bf a}_2$ or by ${\bf a}_3$ and ${\bf a}_1$. 
We note that similar arguments were done for the hyperhoneycomb case in Ref.~\cite{mandal2014}.

\section{Volume constraint}
\begin{figure}[!htb]
  \centering
  \includegraphics[width=\columnwidth,trim = 0 0 0 0,clip]{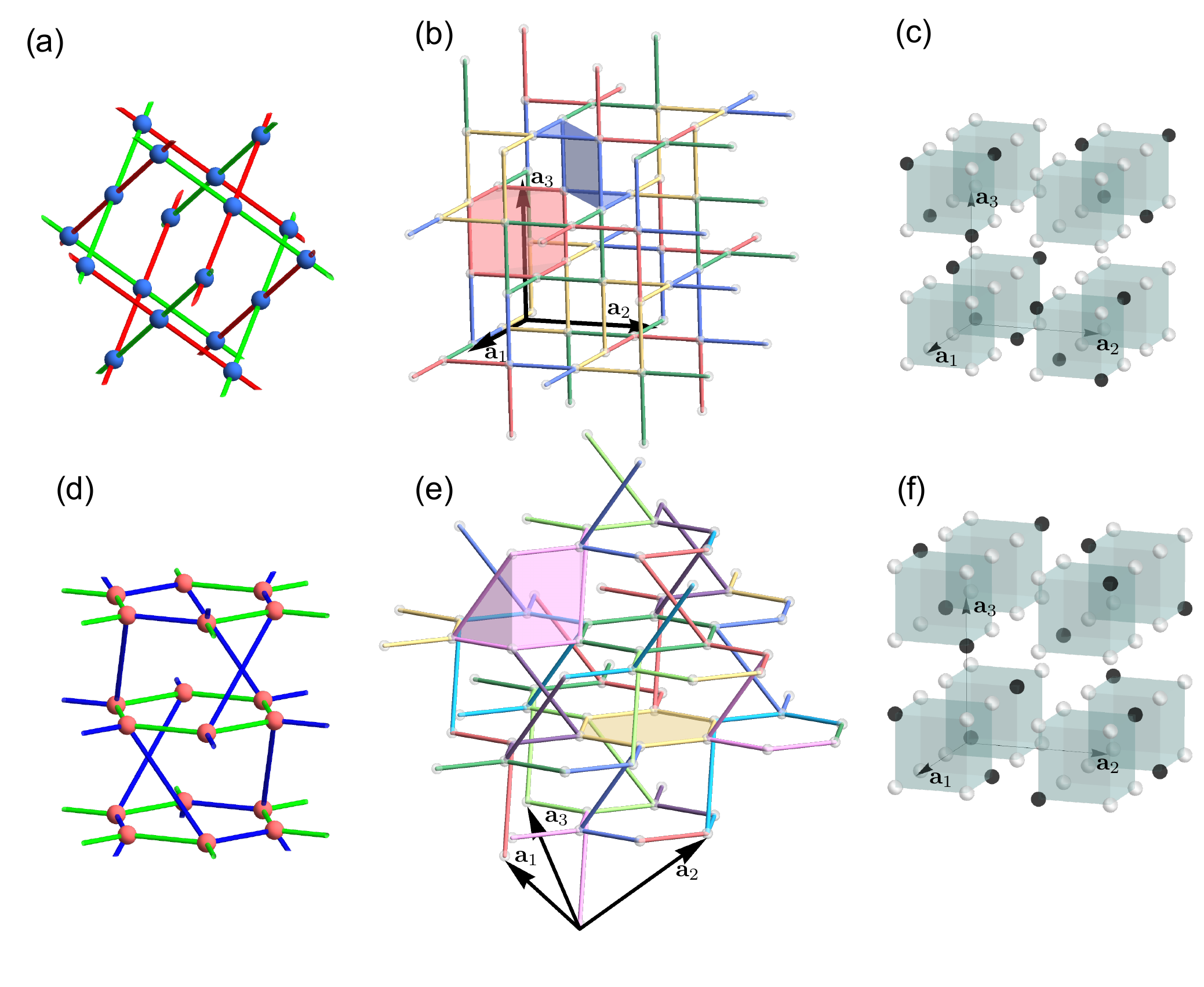}
  \caption{
  	Visualization of the volume constraints for (a-c) the large-$J_z$ limit and (d-f) the large-$J_x$ limit.
	(a)[(d)] 
	Contracted lattice whose sites represent the strong $z$($x$) bonds of the original hypernonagon Kitaev model~[Fig.~\ref{lattice}(a)].
	(b)[(e)] An example of plaquette covering on the contracted lattice for the large $J_z$ ($J_x$) limit.
	A couple of $b_p$ plaquettes are indicated by membranes as examples.  
	(c)[(f)] $b_p$ configuration for the plaquette covering in (b)[(e)]: the product of all the black $b_p$ is unity [Eq.~(\ref{eq:V_constraint})].
  }
  \label{volume}
\end{figure}

Finally, we consider another global ``volume'' constraint
associated with a macroscopic number of fluxes. 
This appears only in the limit of anisotropic interactions. 
In the hypernonagon case in Eq.~(\ref{eq:Kitaev_Hamiltonian}), the limits of $J_z \gg J_x, J_y$ and $J_x \gg J_y, J_z$ give different low-energy effective Hamiltonians 
(the limit of $J_y \gg J_z, J_x$ is equivalent to the latter from the symmetry)~\cite{kato2017}. 
The effective Hamiltonians in the large-$J_z$ and large-$J_x$ limits 
were derived by the perturbation theory as
\begin{eqnarray}
	\mathcal{H}_{\rm eff}^z &=& J \sum_{\langle p,p' \rangle} b_p b_{p'} - J' \sum_{( p,p' )} b_p b_{p'}, \label{eq:heffz}\\
	\mathcal{H}_{\rm eff}^x &=& J_4 \sum_{\langle p_1,p_2,p_3,p_4 \rangle} b_{p_1} b_{p_2}b_{p_3}b_{p_4} - J_2 \sum_{( p,p' )} b_p b_{p'},\label{eq:heffx}
\end{eqnarray}
respectively, with
\begin{equation}
J=\frac{33}{2048} \frac{J_x^4 J_y^4}{|J_z^7|}, \quad
J'= \frac{9}{33} J, \quad
J_2=\frac{9}{2048} \frac{J_y^4 J_z^4}{|J_x^7|}, \quad
J_4= \frac{63}{512} \frac{J_y^6}{|J_x^5|}.
\end{equation}
$b_p$ is the projection of $W_p$ to the low-energy sector, 
which is a $Z_2$ variable taking $\pm 1$.
The sums of $\langle p,p' \rangle$ and $(p,p')$ run over particular bonds on the cubic lattice,
while the sum of $\langle p_1,p_2,p_3,p_4\rangle$ runs over particular sets of four $b_p$ 
(see Ref.~\cite{kato2017} for the details).
For the large $J_\mu$ limit ($\mu = z, x$), we consider a ``contracted" lattice where each $\mu$ bond is contracted to a site located at the center of the bond 
[see Figs.~\ref{volume}(a) and \ref{volume}(d)]. 
Then, the nine-site elementary loops on the original hypernonagon lattice
 become a six-site one on the contracted lattice.
For each six-site loop, we define a $Z_2$ variable $b_p$ as the projection of $W_p$ onto the low-energy sector.
Similar to $W_p$, $b_p$ obeys the surface constraints as well as two local constraints per unit cell: 
the product of all $b_p$ on a surface is unity, and the product of eight $b_p$ in each cube is also unity. 

The volume constraint mentioned above appears as an additional global constraint as follows.
A similar volume constraint was discussed for the hyperhoneycomb case~\cite{mandal2014}.
Following Ref.~\cite{mandal2014}, we consider a covering of all the bonds by $b_p$ plaquettes that uses all the bonds only once on each contracted lattice.
Figures~\ref{volume}(b), \ref{volume}(c), \ref{volume}(e), and \ref{volume}(f) show examples of such coverings $\mathcal{V}$ in each specified limit.
The volume constraint is applied to all the $b_p$ 
on the covering $\mathcal{V}$ as
\begin{eqnarray}
\prod_{p \in \mathcal{V} } b_p =1.
\label{eq:V_constraint}
\end{eqnarray}
Although the choice of $\mathcal{V}$ is not unique, it is enough to consider a particular one, as discussed in the hyperhoneycomb case~\cite{mandal2014}.

\section{Summary}
We have discussed the constraints on $Z_2$ fluxes in two distinct anisotropic limits of the hypernonagon Kitaev model 
whose ground state was demonstrated to be a chiral spin liquid in our recent study~\cite{kato2017}.
We showed that there are two kinds of global constraints in addition to the local constraints: the surface and volume constraints, 
similar to the case of the hyperhoneycomb Kitaev model~\cite{mandal2014}.
The constraints put restrictions on what kind of flux excitations are allowed, which have to be taken into account in the Monte Carlo simulation as the simultaneous flip of a pair of four-site loops~\cite{kato2017}.

\section*{Acknowledgements}
This work was supported by JSPS Grant 
No.~26800199, No.~JP15K13533, No.~JP16K17747, No.~JP16H02206, and No.~JP16H00987.

\section*{References}

\bibliography{mybibfile}

\begin{thebibliography}{10}
\expandafter\ifx\csname url\endcsname\relax
  \def\url#1{\texttt{#1}}\fi
\expandafter\ifx\csname urlprefix\endcsname\relax\def\urlprefix{URL }\fi
\expandafter\ifx\csname href\endcsname\relax
  \def\href#1#2{#2} \def\path#1{#1}\fi

\bibitem{kitaev2006}
A.~Kitaev,
  \href{http://www.sciencedirect.com/science/article/pii/S0003491605002381}{Anyons
  in an exactly solved model and beyond}, Ann. Phys. 321~(1) (2006) 2 -- 111.
\newblock \href {http://dx.doi.org/10.1016/j.aop.2005.10.005}
  {\path{doi:10.1016/j.aop.2005.10.005}}.
\newline\urlprefix\url{http://www.sciencedirect.com/science/article/pii/S0003491605002381}

\bibitem{yao2007}
H.~Yao, S.~A. Kivelson,
  \href{https://link.aps.org/doi/10.1103/PhysRevLett.99.247203}{Exact chiral
  spin liquid with non-abelian anyons}, Phys. Rev. Lett. 99 (2007) 247203.
\newblock \href {http://dx.doi.org/10.1103/PhysRevLett.99.247203}
  {\path{doi:10.1103/PhysRevLett.99.247203}}.
\newline\urlprefix\url{https://link.aps.org/doi/10.1103/PhysRevLett.99.247203}

\bibitem{si2008}
T.~Si, Y.~Yu,
  \href{http://www.sciencedirect.com/science/article/pii/S0550321308003404}{Anyonic
  loops in three-dimensional spin liquid and chiral spin liquid}, Nucl. Phys. B
  803~(3) (2008) 428 -- 449.
\newblock \href
  {http://dx.doi.org/https://doi.org/10.1016/j.nuclphysb.2008.06.009}
  {\path{doi:https://doi.org/10.1016/j.nuclphysb.2008.06.009}}.
\newline\urlprefix\url{http://www.sciencedirect.com/science/article/pii/S0550321308003404}

\bibitem{mandal2009}
S.~Mandal, N.~Surendran,
  \href{https://link.aps.org/doi/10.1103/PhysRevB.79.024426}{Exactly solvable
  kitaev model in three dimensions}, Phys. Rev. B 79 (2009) 024426.
\newblock \href {http://dx.doi.org/10.1103/PhysRevB.79.024426}
  {\path{doi:10.1103/PhysRevB.79.024426}}.
\newline\urlprefix\url{https://link.aps.org/doi/10.1103/PhysRevB.79.024426}

\bibitem{hermanns2015}
M.~Hermanns, K.~O'Brien, S.~Trebst,
  \href{https://link.aps.org/doi/10.1103/PhysRevLett.114.157202}{Weyl spin
  liquids}, Phys. Rev. Lett. 114 (2015) 157202.
\newblock \href {http://dx.doi.org/10.1103/PhysRevLett.114.157202}
  {\path{doi:10.1103/PhysRevLett.114.157202}}.
\newline\urlprefix\url{https://link.aps.org/doi/10.1103/PhysRevLett.114.157202}

\bibitem{obrien2016}
K.~O'Brien, M.~Hermanns, S.~Trebst,
  \href{https://link.aps.org/doi/10.1103/PhysRevB.93.085101}{Classification of
  gapless $z_{2}$ spin liquids in three-dimensional kitaev models}, Phys. Rev.
  B 93 (2016) 085101.
\newblock \href {http://dx.doi.org/10.1103/PhysRevB.93.085101}
  {\path{doi:10.1103/PhysRevB.93.085101}}.
\newline\urlprefix\url{https://link.aps.org/doi/10.1103/PhysRevB.93.085101}

\bibitem{kitaev2003}
A.~Kitaev,
  \href{http://www.sciencedirect.com/science/article/pii/S0003491602000180}{Fault-tolerant
  quantum computation by anyons}, Annals of Physics 303~(1) (2003) 2 -- 30.
\newblock \href
  {http://dx.doi.org/http://dx.doi.org/10.1016/S0003-4916(02)00018-0}
  {\path{doi:http://dx.doi.org/10.1016/S0003-4916(02)00018-0}}.
\newline\urlprefix\url{http://www.sciencedirect.com/science/article/pii/S0003491602000180}

\bibitem{kimchi2014}
I.~Kimchi, J.~G. Analytis, A.~Vishwanath,
  \href{https://link.aps.org/doi/10.1103/PhysRevB.90.205126}{Three-dimensional
  quantum spin liquids in models of harmonic-honeycomb iridates and phase
  diagram in an infinite-$d$ approximation}, Phys. Rev. B 90 (2014) 205126.
\newblock \href {http://dx.doi.org/10.1103/PhysRevB.90.205126}
  {\path{doi:10.1103/PhysRevB.90.205126}}.
\newline\urlprefix\url{https://link.aps.org/doi/10.1103/PhysRevB.90.205126}

\bibitem{mandal2014}
S.~Mandal, N.~Surendran,
  \href{https://link.aps.org/doi/10.1103/PhysRevB.90.104424}{Fermions and
  nontrivial loop-braiding in a three-dimensional toric code}, Phys. Rev. B 90
  (2014) 104424.
\newblock \href {http://dx.doi.org/10.1103/PhysRevB.90.104424}
  {\path{doi:10.1103/PhysRevB.90.104424}}.
\newline\urlprefix\url{https://link.aps.org/doi/10.1103/PhysRevB.90.104424}

\bibitem{wells1977}
A.~F. Wells, Three dimensional nets and polyhedra, Wiley, New York, 1977.

\bibitem{kato2017}
Y.~Kato, Y.~Kamiya, J.~Nasu, Y.~Motome, Chiral spin liquids at finite
  temperature in a three-dimensional kitaev model, arXiv:1707.02794.

\end{thebibliography}

\end{document}